\documentclass[preprint,fleqn]{revtex4}
\usepackage{bm}

\usepackage{amsmath}
\usepackage{float}
\usepackage{dcolumn}
\usepackage{amssymb}
\usepackage{graphics}
\usepackage{amsmath, amsfonts, amsthm, amssymb, graphicx,comment}
\usepackage{etex}

\newcommand{\be}{\begin{equation}}
\newcommand{\ee}{\end{equation}}
\newcommand{\ba}{\begin{align}}
\newcommand{\ea}{\end{align}}
\newcommand{\n}{\nonumber}
\newcommand{\bea}{\begin{eqnarray}}
\newcommand{\eea}{\end{eqnarray}}

\begin{document}

\title{Cosmological evolutions of $F(R)$ nonlinear massive gravity }

\author{De-Jun Wu\footnote{Email: wudejun10@mails.ucas.ac.cn}}
\affiliation{School of Physics, University of Chinese Academy of Sciences, Beijing 100049, China}

\begin{abstract} Recently a new extended nonlinear massive gravity model   has  been proposed which includes the $F(R)$ modifications to dRGT model.
We follow the $F(R)$ nonlinear massive gravity and study its
implications  on cosmological evolutions. We derive the critical
points of the cosmic system and   study the corresponding kinetics
by performing the  phase-plane analysis.

 \end{abstract}

\maketitle

\section{INTRODUCTION}
\maketitle

The search for a consistent  covariant modification of General
Relativity in which  graviton is allowed to acquire a mass has
been initiated since Fierz and Pauli  (FP) proposed a quadratic
Lagrangian which describes a massive spin-2
field\cite{Fierz:1939ix}. The Lagrangian is ghost-free
\cite{VanNieuwenhuizen:1973fi} but can not recover  Einstein
gravity in the limit of vanishing graviton mass, due to the
coupling between the longitude mode of the graviton and the trace
of the energy momentum tensor\cite{vanDam:1970vg,
Zakharov:1970cc}. Nonlinear terms were introduced and the
troublesome mode could be suppressed at macroscopic length scales
via the Vainshtein mechanism\cite{Vainshtein:1972sx}. However the
same nonlinear terms are responsible for the existence of the
Boulware-Deser (BD) ghost\cite{Boulware:1973my,
ArkaniHamed:2002sp, Creminelli:2005qk, Deffayet:2005ys,
Gabadadze:2003jq} which would make the theory unstable. Until
recent, a scheme was invented by de Rham, Gabadadze and
Tolley(dRGT)  so that only the suitable nonlinear terms enter the
theory thus eliminating the BD ghost once for all
\cite{deRham:2010ik, deRham:2010kj, Hassan:2011vm, Hassan:2011hr,
deRham:2011rn, deRham:2011qq} (see \cite{Hinterbichler:2011tt,
deRham:2014zqa} for  review),  implications of the dRGT model has
been studied in \cite{D'Amico:2011jj, Gumrukcuoglu:2011ew,
Gumrukcuoglu:2011zh,  Gumrukcuoglu:2012aa, Koyama:2011wx,
Comelli:2011zm, Crisostomi:2012db, Cardone:2012qq, Gratia:2012wt,
Kobayashi:2012fz, D'Amico:2012pi, Fasiello:2012rw,
Langlois:2012hk, Gong:2012yv,  Saridakis:2012jy, Cai:2012ag,
Cai:2012db, Rham:2010ri, Akrami:2012ri, Akrami:2013pna,
Volkov:2013roa, Li:2013gpa, DeFelice:2013bxa,  Tasinato:2013rza,
D'Amico:2013kya,  Khosravi:2013axa, Langlois:2013cya,
Akrami:2013ffa, Andrews:2013uca,
Maggiore:2013mea,Tamanini:2013xia,Comelli:2013tja,Fasiello:2013woa,
Blake:2013bqa, Bamba:2013fha, Bamba:2013aca, Bamba:2013hza,
Foffa:2013vma, Kobayashi:2014mpa, Nesseris:2014mea, Goon:2014ywa,
Tasinato:2014eka}.

Some extended nonlinear massive gravity theories were introduced
shortly after. In mass-varying massive gravity theory
\cite{Huang:2012pe}, the graviton mass is promoted to vary with a
dynamical scalar field, the cosmological evolutions of the model
have been studied in \cite{Wu:2013ii, Hinterbichler:2013dv,
Leon:2013qh, DeFelice:2013awa, Gumrukcuoglu:2013nza,
Huang:2013mha}. Especially in \cite{Cai:2012ag}, bounce and cyclic
cosmology has been builded, which might have interesting
implications \cite{Piao:2004me, Piao:2009ku, Piao:2010py,
Zhang:2010bb} and help to confronting the massive gravity with
observation, e.g. explaining CMB anomalies \cite{Piao:2003zm,
Piao:2005py, Liu:2013kea}. In quasi-dilaton theory
\cite{D'Amico:2012zv}, a dynamical scalar field is also present
but it is  non-trivial coupling with the massive graviton instead,
see also \cite{Haghani:2013eya, Gannouji:2013rwa}. The extended
theories have more theoretical freedom and thus allow for more
desirable cosmological solutions \cite{DeFelice:2013dua}.

Recently a new extended theory was proposed which introduce the
$F(R)$ modifications to the dRGT model\cite{Cai:2013lqa}. The
theory contains   modification of GR not only in IR regimes like
all the other massive gravity theories but also
 in UV regimes. The theory is free of the DB ghost as proven
in\cite{Cai:2013lqa}, and inherit the theoretical  advantages of $F(R)$ paradigm\cite{Cai:2014upa}. Later it is also
claimed to be free of ghosts instabilities at perturbative level which was
found in dRGT model\cite{DeFelice:2012mx}. The theory
allows a huge class of interesting cosmological behaviors at early and late times, and is promising in fitting the current observations. Ghost-free $F(R)$
bigravity was proposed in \cite{Nojiri:2012zu} which would give variety of cosmic acceleration models\cite{Nojiri:2012re}. A different model of ghost-free massive $F(R)$ gravity was develop in \cite{Kluson:2013yaa}.

In the paper we analyze the $F(R)$ nonlinear massive gravity  model and study  its  implications  on cosmic evolution by performing
a phase-space and stability analysis. The  dynamical analysis  of the regular
$F(R)$ gravity has been carried out  in some papers \cite{Amendola:2006we, Carloni:2007br, Leon:2013bra}, for general analyses see \cite{Nojiri:2010wj, Nojiri:2006ri}.  $F(G)$  model
has been analyzed in \cite{Zhou:2009cy}, see also \cite{Nojiri:2005jg}. However in the present model the dynamic of the
system becomes  more complicated,  not only the dimensionless variables have more complex evolution, but some of the variables are
not completely independent, and their relations yield extra constraints on the
system which make the system  more difficult to study. In the previous
paper \cite{Amendola:2006we} a technique was developed which enables one to analyze the dynamic of
the system without a specific model of $F(R)$, we will see that this technique is partially valid in the present work. The stability  of  fixed points
has a strong correlation with the specific model of $F(R)$, so we will
analyze two  models after giving a general discussion of solution space
of the system.

The present paper is organized as follows. In Section II, we briefly review  the $F(R)$ nonlinear massive gravity  model and its cosmological equations of motion. Then we construct the dynamics of $F(R)$ nonlinear massive gravity and develop a method to deal with a system with extra constraints in Section III. We  provide the solutions of the cosmic system described by this theory by performing  detailed phase-space and stability analyses of two $F(R)$ models and summarize the results in Section IV. Finally, we conclude with a discussion in Section V.

\section{Cosmology of $F(R)$ nonlinear massive gravity  }

To begin with, we briefly review the $F(R)$ nonlinear massive gravity model constructed in \cite{Cai:2013lqa}. This model  imposes a UV sector modification
of the dRGT model with the scalar curvature $R$ replaced by an  arbitrary
function of it.
Therefore, the complete action can be expressed as
\begin{eqnarray}\label{action}
 S = \frac{M_p^2}{2} \int d^4x \sqrt{|g|} ~ [ F(R) + 2m_g^2 {\cal U}_M ] ,
\end{eqnarray}
where $M_p$ is the Planck mass, $g$ is the physical metric and $m_g$ is the graviton
mass. Graviton potential is given by ${\cal U}_M = {\cal U}_2 +
\alpha_3 {\cal U}_2 + \alpha_4 {\cal U}_2 $, where $\alpha_3$, $\alpha_4$ are  dimensionless parameters, and
\begin{eqnarray}\label{UMs}
 {\cal U}_2 = \mathcal{K}^\mu_{[\mu} \mathcal{K}^\nu_{\nu]},~
 {\cal U}_3 = \mathcal{K}^\mu_{[\mu} \mathcal{K}^\nu_{\nu}
\mathcal{K}^\sigma_{\sigma]},~
 {\cal U}_4 = \mathcal{K}^\mu_{[\mu} \mathcal{K}^\nu_{\nu}
\mathcal{K}^\sigma_{\sigma} \mathcal{K}^\rho_{\rho]},
\end{eqnarray}
where  $\mathcal{K}
\equiv \mathcal{I} - \sqrt{g^{-1}f}$, and $f$ denote the fiducial metric.
We begin  with a Minkowski fiducial metric
\be
 f_{AB}=\eta_{AB},
\ee
and  an  open FRW physical metric
\begin{eqnarray}
ds^2=-N^2dt^2+a^2(t) \gamma^{K}_{ij}dx^idx^j,
\end{eqnarray}
where
\be\gamma^{K}_{ij} dx^idx^j = \delta_{ij}dx^idx^j -
\frac{a_0^2(\delta_{ij}x^idx^j)^2}{1+a_0^2\delta_{ij}x^ix^j}\n\ee
and
$a_0=\sqrt{|K|}$, $a_0$  is associated with the spatial curvature.
Variation of the action with respect to $b$, $N$ and $a$ gives  three equations
\begin{align}
\label{constraint}
 &(\dot{a}-a_0)Y_1 = 0, \\
 \label{em1}
 &3 M_p^{2} F_{,R} \left( H^2-\frac{a_0^2}{a^2} \right) = \rho_m+\rho_{\rm
IR}+\rho_{\rm UV},\\
 \label{em2}
 &M_p^{2} F_{,R}\left(-2\dot{H}-3H^2 +\frac{a_0^2}{a^2}\right) = p_m+p_{\rm
IR}+p_{\rm UV},
\end{align}
where $\dot{a}=\frac{a'}{N}$ and $H=\frac{\dot{a}}{a}$. In the above expressions
 IR (massive gravity) and UV ($F(R)$ sector) effective contributions   are defined as follow
 \begin{align}
 & \rho_{\rm IR} = m_g^2M_p^2({\cal B}-1)(Y_1+Y_2), \\
 & p_{\rm IR} = -m_g^2M_p^2({\cal B}-1)Y_2-m_g^2M_p^2(\dot{b}-1)Y_1,\\
 & \rho_{\rm UV} = M_p^2 \left[\frac{R F_{,R}-F}{2}-3H\dot{R} F_{,RR}\right],
\\
\label{PfR}
 & p_{\rm UV} = M_p^2 \left[\dot{R}^2 F_{,RRR}+2H\dot{R} F_{,RR}+\ddot{R}
F_{,RR} +\frac{F-RF_{,R}}{2} \right],
\end{align}
where the polynomials $Y_{1,2}$ are given by
$Y_1 = (3-2{\cal B})+\alpha_3(3-{\cal B})(1-{\cal B})+\alpha_4(1-{\cal B})^2$
and
$Y_2 = (3-{\cal B})+\alpha_3(1-{\cal B})$ with ${\cal B}=\frac{a_0b}{a}$.
Similar to all massive gravity scenarios, the nontrivial solutions of Eq. \eqref{constraint} correspond to the case of $Y_1=0$ and yield
\begin{align}
 {\cal{B}}_\pm = \frac{1 +2\alpha_3 -2\alpha_4\pm\sqrt{1 +\alpha_3
+\alpha_3^2
-\alpha_4}}{\alpha_3 +\alpha_4}.
\end{align}
This relation can be fulfilled by choosing $b(t)\propto a(t)$, and
therefore  it yields $\rho_{\rm IR}=-p_{\rm IR}$ to be constant.

\section{DYNAMICAL FRAMEWORK }

In this section we perform a detailed phase-space analysis of cosmic evolutions describing the $F(R)$ nonlinear massive gravity model following the method  developed in \cite{Halliwell:1986ja, Ferreira:1997au, Copeland:1997et, Guo:200304, Guo:200306, Guo:2004fq, Gong:2006sp, Cai:2006dm, Copeland:2006wr, Zhou:2007, Chen:2008ft, Zhou:2009cy, Cai:2009zp} (see also \cite{Leon:2012mt} for a recent analysis in the frame of generalized Galileon cosmology).
First we give a brief review of autonomous system and  transform the dynamical system into the autonomous form,
 then discuss a problem  in the phase-space analysis and
try to give a possible  solution.

\subsection{Autonomous system and its evolution}
For a general dynamical system, one  group of suitable auxiliary variables  can be chosen so that
 the corresponding equations of motion will be first-order differential equations.
The group of auxiliary variables can be written as a vector  $\vec{x} $,
its equation of motion is
\begin{align}
\frac{d\vec{x}}{d t} = \vec{f}(x).
\end{align}
The system is said to be autonomous if $\vec{f }(x)$ do not contain explicit time-dependent terms. We want to find out which state will the system
be, eventually. If the system is stabilized at one particular
state, the speed of the variables must equal to $0$, assuming the number of  variables  is $n$, this condition corresponds to:
\begin{align}
\left\{ {\begin{array}{*{20}c}
   {f_1 \left( {x_1 {\rm{,}}x_2 {\rm{,}} \cdots x_n } \right) = 0}  \\
    \vdots   \\
   {f_n \left( {x_1 {\rm{,}}x_2 {\rm{,}} \cdots x_n } \right) = 0}.\\
\end{array}} \right.
\end{align}
The solutions to these equations are called fixed points, they are the
candidates for stable states. In order to find out whether a solution is stable
or not, we need to analyze the perturbation around it.
 By taking the perturbation of the system  we get
\begin{align}
\label{eqa}
\frac{{d\delta \vec x}}{{dt}} = A \cdot \delta \vec x,
\end{align}
where $A$ is a matrix with element  $
A_{ij}  = \frac{{\partial f_i }}{{\partial x_j }}\left( {\vec x_0 } \right)
$, $\vec x_0$ is the fixed point under study. We can view the equation above as the equation of motion of the
perturbation.
Assuming eigenvalues of $A$ are $\left( {\mu _1  \cdots \mu _n } \right)$, and the corresponding eigenvectors are
$\left( {\vec{\nu}} _1  \cdots {\vec{\nu}} _n \right)$, if the perturbation starts
as  $ {\vec{\nu}} _m$, then its evolution takes the form of ${\vec{\nu}} _m \exp \left[ {\mu_m t} \right]$. Because the eigenvectors of $A$ are
linearly independent in most cases, any perturbation can be written as $\delta \vec{x}\left( {t = 0} \right) = \sum\limits_{i = 1}^n {\alpha _i \vec{\nu} _i }$, where $\alpha _i$ is arbitrary coefficient, and because Eq.\eqref{eqa} is linear,   the evolution of the perturbation follows the equation \be
\delta \vec x(t) = \sum\limits_{i = 1}^n {\alpha _i {\vec{\nu}} _i } e^{\mu _i t}.\ee If the the perturbation around a fixed point  becomes smaller over time and approaches $0$ eventually, that is $
\mathop {\lim }\limits_{t \to  + \infty } \delta \vec x = 0
$, we call this  fixed  point  asymptotic stable. One can see that this requirement
is satisfied when the real parts of all the eigenvalues are negative.
In cosmology, we mainly concern the stable fixed points, for they contain information about  late-time evolution of the universe.

\subsection{Dynamics of $F(R)$ nonlinear massive gravity}
We begin to discuss the dynamic system of the $F(R)$ nonlinear massive gravity
in detail. First we define $6$ dimensionless variables
\begin{align}
&x_1  = \frac{{\rho _{IR} }}{{3M_p^2 F_{,R} H^2 }},
x_2  = \frac{R}{{6H^2 }},
x_3  =  - \frac{F}{{6F_{,R }H^2 }},
x_4  =  - \frac{{{\dot R}F_{,RR} }}{{F_{,R} H}},\n\\
&x_5  = \frac{{a_0^2 }}{{a^2 H^2 }},
\Omega _m  = \frac{{\rho _m }}{{3M_p^2 F_{,R} H^2 }}.\n
\end{align}

Eq.\eqref{em1} can be reduced to
\begin{align}
\label{xianzhi}\Omega _m  + x_1  + x_2  + x_3  + x_4  + x_5  = 1.
\end{align}
Using Eq.\eqref{em1} and Eq.\eqref{em2}, we can write the equations of motion of the variables  as follow
\begin{align}
\label{x1}&\frac{d}{{dN}}x_1  = x_1 \left( {x_4  - 2\left( {x_2  + x_5 } \right) + 4} \right),\\
\label{x2}&\frac{d}{{dN}}x_2  = x_2 \left( { 4- 2\left( {x_2  + x_5 } \right) } \right) - \frac{{x_2 x_4 }}{m},\\
\label{x3}&\frac{d}{{dN}}x_3  = x_3 \left( {x_4  - 2\left( {x_2  + x_5 } \right) + 4} \right) + \frac{{x_2 x_4 }}{m},\\
\label{x4}&\frac{d}{{dN}}x_4  = x_4 \left( {x_4  - x_2  - x_5 } \right) + x_5  - x_2  - 3\left( {x_1  + x_3 } \right) + 3\omega \Omega _m  - 1,\\
\label{x5}&\frac{d}{{dN}}x_5  = 2x_5 \left( {1 - \left( {x_2  + x_5 } \right)} \right),\\
\label{x6}&\frac{d}{{dN}}\Omega _m  = \Omega _m \left( x_4  - 2\left( {x_2  + x_5 } \right) + 1 - 3{\omega } \right),
\end{align}
where $N$ stands for $lna$, and $m = \frac{{F_{,RR} R}}{{F_{,R} }}$, it's a parameter that depends on the form of $F(R)$.
 We can define another parameter $r = \frac{{x_2 }}{{x_3 }} =  - \frac{{F_{,R} R}}{F}$, once we have the exact form of  $F(R)$, we can derive $R$ from $r$
and then substituting it into $m$, in the end the parameter $m$
will be a function of $x_2$, $x_3 $ and the dynamical system will  become autonomous.

 Eq.\eqref{xianzhi} must hold at anytime, differentiating it respect to $N$ gives
\be
\frac{d}{{dN}}(x_1  +x_2  +x_3  +x_4  +x_5  + \Omega _m)=0.
\ee
Adding
Eq. \eqref{x1} through  Eq. \eqref{x6} we get
\begin{align}
&\frac{d}{{dN}}(x_1  +x_2  +x_3  +x_4  +x_5  + \Omega _m)\n\\
&=(1-2 (x_2+2 {x_5})+x_4) (\Omega_m+{x_1}+{x_2}+{x_3}+{x_4}+{x_5}-1)\label{zhengjiao}.
\end{align}
One can see from Eq.\eqref{zhengjiao} that the constraint equation Eq.\eqref{xianzhi}
is not automatic  satisfied,  we must impose it when we solve the
system, and after doing so we can eliminate one variable,
 we eliminate $\Omega_m$ for convenience
and do not consider Eq.\eqref{x6}.

From the definitions of $x_1,~x_2 $ and $x_3$, we can express $R$ and $H$ in terms of any two variables and substitute them into the third one,
which allows us  to eliminate one variable directly,  thus reducing the dimension of the dynamical system by $1$. However for some
complicated  expression of  $F(R)$,  it is not always possible to give the resolved expression of $R$ and $H$, and  in some cases  eliminating one variable  would  actually make the dynamics of system more complicated because the complex relations between $x_1,~x_2 $ and $x_3$ will be involved in the equations of motion.   If we do not consider the relations between the variables, we could get
false result, unnecessary fixed points may appear, and
the analysis from the whole phase space may be  unreliable.
We will discuss this issue in the next section.

\subsection{System with hidden constraint   }

We now consider the issue  that one  must compute the fixed points
and study the stability of these fixed points with redundant
degrees of freedom in the system.
 $x_1$, $x_2$ and $x_3$ are not entirely
independent, and the system is restricted to a low-dimensional surfaces.
The surface is uniquely determined, since $x_1$, $x_2$ and $x_3$ are all
functions of $R$ and $H$, $R$ and $H$ are the natural coordinates of the surface, we will call this surface $h(x_1, x_2, x_3)=0$.

$x_1$, $x_2$ and $x_3$  do not contain  contain explicit time-dependent terms, neither does the  surface  $h=0$, thus
 all the orbits of the system are guaranteed to stay on the surface  and  the speed of the system is  tangent to
the surface at any time, the condition is characterized by the expression
\be
\sum\limits_n^{} {\frac{{\partial h}}{{\partial x_n }}\frac{{dx_n }}{{dN}} = 0},
\ee
which implies that $h=c$  is an integral, where $c$ is a constant and  $n$ is the number of  variables. But the actual system only stays on the surface of
$h=0$. In the previous section, we have another constraint equation
\begin{align}
\Omega _m  + x_1  + x_2  + x_3  + x_4  + x_5  = 1.
\end{align}
It is low-dimensional plane with  normal vector $(1, 1, 1, 1, 1, 1)$, and Eq.\eqref{zhengjiao} prove that it is an integral.
Usually finding the integral
of a dynamic system is a difficult task, but we know the system under study
must have integrals, because the precise forms of the variables are given.
It seems straightforward that one should  include equation $h=0$ when
solving the system for fixed
points and analyzing the  perturbations of the system around these  fixed points, reflecting  this fact is that all the fixed points must stay on the constraint surface and the  perturbation of the system
belongs to the tangent plane of the given  point,
\be
\sum\limits_n {\frac{{\partial h}}{{\partial x_n }}\delta x_n }  = 0.
\ee
A natural question is that will the perturbation of the system stay in
the same tangent plane as it  evolves.
One can see that this condition is not always satisfied  by taking the
time derivative of the  equation above,
\be
\sum\limits_n {\left( {\frac{{\partial h}}{{\partial x_n }}\frac{{d\delta x_n }}{{dN}} + \sum\limits_m {\frac{{\partial ^2 h}}{{\partial x_m \partial x_n }}\frac{{dx_m }}{{dN}}\delta x_n } } \right)}  = 0.
\ee
If $h$ is a linear function, then $ \frac{{\partial ^2 h}}{{\partial x_m \partial x_n }}$ equals $0$, equation above becomes $\sum\limits_n  {\frac{{\partial h}}{{\partial x_n }}\frac{{d\delta x_n }}{{dN}}}=0$,
which means that at any point the speed of the perturbation always stays in  the tangent plane,  so does the  perturbation.
The constraint surface we encounter before is linear function.
But for  general  surfaces
$ \frac{{\partial ^2 h}}{{\partial x_m \partial x_n }}\neq0$,
in this case the speed of  the perturbation will  no longer stay in the tangent plane beacuse
 $\sum\limits_n  {\frac{{\partial h}}{{\partial x_n }}\frac{{d\delta x_n }}{{dN}}}\neq0$, neither will the perturbation.
But we will see that this fact won't  cause trouble because
 we  only consider the  perturbations around fixed points,
and at the fixed point we have  $\frac{{dx_m }}{{dN}}=0$,  thus
$ \sum\limits_n  {\frac{{\partial h}}{{\partial x_n }}\frac{{d\delta x_n }}{{dN}}}=0$ still holds, this result could simplify the calculation. We could assume that the perturbations
start on the  tangent plane and  studying their evolutions  without concerning  their unwanted behaviors.

The perturbation   around any given fixed point could be
written  as \be
\delta \vec{x} = \sum\limits_{i = 1}^{n-1} {\beta _i \vec{e}_i }  = \sum\limits_{j = 1}^n {\alpha_j } \vec{\nu} _j,\ee where $\vec{e}_i$ is the   base vector of the tangent plane, $\vec{\nu} _j$ is  the eigenvector of the system, because the eigenvectors of a certain point are   linearly independent, $\vec{e}_i$ and $\vec{n}$ (the normal vector of the surface) can be written as the linear combinations of these eigenvectors, then the  perturbation can be written  as combination of $\vec{\nu} _j$, $\alpha _j$ is the coefficient.
When a fixed point is stable viewing from the  whole phase   space, the stable subspace is $n$ dimensional, tangent plane  must belong to  it, and the
fixed point is stable because  the perturbation will approach $0$ as time passing by regardless of the actual value of the  coefficient $\beta _i$. If the stable subspace is $n-1$ dimensional, there is a chance that the tangent plane belongs to it  and the point is still stable.

Generally speaking, if  $\alpha_j$ equals $0$, the corresponding eigenvalue do not effect the stability of the fixed point. This happens when    the eigenvector is normal to the surface or the other  eigenvectors
belong to the tangent plane. After we find out which eigenvalue is responsible for the stability of the fixed point, we could give the parameter range
for the point to be stable.

\section{Phase-space analysis and results}
 A special method has been developed for  analyzing $F(R)$ model which enables one to obtain a general understanding of the system
without specific form of $F(R)$\cite{Amendola:2006we}.
Instead of  solving the specific form of $m(r)$, one could  solve the system for fixed points assuming $m$ is another unknown variable. Some fixed points do not
contain $m$ and are assumed to be independent  of form of $F(R)$, the rest fixed points contain $m$ and can give a new relation between $m$ and $r$ because by definition $r$ is the ratio of  $x_2$ and $x_3$,  both of which contain $m$. Together with the function of $m(r)$ imply by the $F(R)$, one can find the exact  value of $r$ and $m$, and the fixed points are  determined. One can even carry out the stability analysis without knowing the form of
$m(r)$ and the value of $r$, although in \cite{Carloni:2007br}, the authors claim that the  stability analysis in this way can be troublesome, and because the form of $F(R)$ strongly influences the stability of these fixed points in the $F(R)$ nonlinear massive gravity, we will not carry out the  stability analysis in this way, instead we will study  two models and give detailed analysis of each fixed point.

The system may have less fixed points due to the constraints of the system, but all the possible fixed points can be
found, after that  finding the fixed points existed in a certain  $F(R)$ model  is relatively easy, we just keep the fixed points that satisfy the constraint. All the possible fixed points are listed
\begin{align}
&A:\left( {5 - x_3, 0, x_3, - 4, 0} \right)\n\\
&B:\left( {2 - x_3, 0, x_3, - 2, 1} \right)\n\\
&C:\left( { - 1 - x_3, 2, x_3, 0, 0} \right)\n\\
&D:\left( {0, 0, 0, 0, 1} \right)\n\\
&E:\left( {0, 0, 0, 1, 0} \right)\n\\
&F:\left( {0, 0, 0, - 1 + 3\omega, 0} \right)\n\\
&G:\left( {0, 0, 0, 1 + 3\omega, 1} \right)\n\\
&H:\left( {0, - 1 - 3\omega, 2 + 6\omega, - 3\left( {1 + \omega } \right), 0} \right)m \to  - \frac{1}{2}\n\\
&I:\left( {0, - \frac{3}{2}x_3 \left( {1 + \omega } \right), x_3, 1 + 3\omega ,\frac{1}{2}\left( {2 + 3x_3 \left( {1 + \omega } \right)} \right)} \right)m \to \frac{1}{2}\left( {1 + 3\omega } \right)\n\\
&J:\left( {0, 2m(1 + m), - 2m, 2m, 1 - 2m - 2m^2 } \right)\n\\
&K:\left( {0, \frac{{4m^2+3m    - 1  }}{{m(1 + 2m)}}, \frac{{1 - 4m}}{{m + 2m^2 }}, \frac{{2 - 2m}}{{1 + 2m}}, 0} \right)\n\\
&L:\left( {0, \frac{{1 + 4m - 3\omega }}{{2 + 2m}}, \frac{{3\omega - 1 - 4m  }}{{2(1 + m)^2 }}, \frac{{3m(1 + \omega )}}{{1 + m}}, 0} \right)\n.
\end{align}
From points I, J, K, L we can derive the relation
\be
m\left( r \right) =  - 1 - r.
\ee
For each $F(R)$ model one or more values of  $r$ can  be determined,
and the exact form of fixed points can be obtained.
Notice that A, B, C, I are  lines of equilibria instead of fixed points,   we will
not be bothered  by this fact because when considering the system with
a constraint surface,  the lines of equilibria will intersect with it and the
points of intersection are the fixed points of the system.

 The stability of the fixed points is related to  the
constraint surface and can not be analyzed  without the precise form of $F(R)$, we will consider two  $F(R)$ models and  analyze the stability of each fixed point.
\subsection{$R^n$ model}

Let us consider the Lagrangian
$
F\left( R \right) = R^n
$.
Corresponding constraint surface can be written as
\be
n x_3  + x_2  = 0.
\ee
First we check if the surface is an  integral, after some calculation we get
\be
\vec{n}\cdot\frac{d\vec{x}}{dN}=(n x_3+{x_2}) (2 {x_2}+4-2 ( {x_5}+{x_4})),
\ee
so the surface is an  integral, the normal vector of the surface is $\vec{n}=\left( {0, 1, n, 0, 0} \right)$, $m$ takes the constant value of  $n-1$. We summarize the fixed points of this autonomous system and their stability  in
Table I and Table II, respectively.
\begin{table}
\begin{tabular}{ccccccc}

\hline
Lable &${x_1}$ & ${x_2}$ & ${x_3}$ & ${x_4}$ & ${x_5}$ & ${\Omega_m}$ \\
\hline$A_1$&5&0&0&-4&0&0\\
$B_1$&2&0&0&-2&1&0\\
$C_1$&$- 1 + \frac{2}{n}$&$2$& $- \frac{2}{n}$&0&0&0\\
$D_1$&0&0&0&0&1&0\\
$E_1$&0&0&0&1&0&0\\
$F_1$&0&0&0&$- 1 + 3\omega $&0&$2-3 \omega$\\
$G_1$&0&0&0&$1 + 3\omega$&1&$-3 \omega -1$\\
$J_1$&0&$2n\left( {n - 1} \right)$&$2\left( {1 - n} \right)$&$2\left( {n - 1} \right)$&$1 + 2n\left( {1 - n} \right)$&$0$\\
$K_1$&0&$\frac{{n( 4n- 5)}}{{(2n-1)(n-1)  }}$&$\frac{{(5 - 4n)}}{{(2n-1)(n-1) }}$&$  \frac{{2\left( {n - 2 } \right)}}{{ 1-2n }}$&$0$&$0$\\
$L_1$&0&$  \frac{{  4n - 3\omega  - 3}}{{2n}}$&$\frac{{ 3\omega - 4n   + 3}}{{2n^2 }}$&$\frac{{3(n - 1)(\omega  + 1)}}{n}$&$0$&$\frac{n (9 \omega +13-2 n (3 \omega +4))-3 (\omega +1)}{2 n^2}$\\
\hline
\end{tabular}
\caption{The fixed points   in $F(R)$ nonlinear massive gravity with $F(R)=R^n$.}
\end{table}

\begin{table}
\begin{tabular}{ccc}

\hline
Point~~~~~~~~~~~~~ &~~~~~~~~~~ $\omega_{eff}$~~~~~~~ &~~~~~~~~ Stable when~~~~~~~~~ \\
\hline$A_1$~~~~~~~~~~~~~ & $\frac{1}{3} $  &  not stable  \\
$B_1$~~~~~~~~~~~~~&  $-\frac{1}{3} $ & $ 0 < n < 1,~\omega  >  - 1$\\
$C_1$~~~~~~~~~~~~~& $-1 $  &$ 1 < n < 2,~ \omega  >  - 1$ \\
$D_1$~~~~~~~~~~~~~&  $-\frac{1}{3} $ & not stable \\
$E_1$~~~~~~~~~~~~~&  $\frac{1}{3} $ &not stable \\
$F_1$~~~~~~~~~~~~~& $\frac{1}{3}$  & not stable\\
$G_1$~~~~~~~~~~~~~& $-\frac{1}{3} $  &${n < 0,~ \omega  < \frac{1}{3}(2n - 3)}
$
or
$
0 < n < 1,~ \omega  <  - 1 $ \\
$J_1$~~~~~~~~~~~~~&  $ -\frac{1}{3}$ & $\frac{1}{2} - \frac{{\sqrt 3 }}{2} < n < 0,~
\omega  > \frac{1}{3}(2n - 3)
 $\\
$K_1$~~~~~~~~~~~~~& $\frac{n (7-6 n)+1}{6 n^2-9 n+3} $  & $n < \frac{1}{2}\left( {1 - \sqrt 3 } \right)
$ ~or~$\frac{1}{2} < n < 1 $~or~$
n > 2 $,
and
$
{ \omega  > \frac{{ - 8n^2  + 13n - 3}}{{6n^2  - 9n + 3}}} $\\
$L_1$~~~~~~~~~~~~~& $\frac{\omega +1}{n}-1 $  &stable \\
\hline
\end{tabular}
\caption{The stability of the fixed points    in $F(R)$ nonlinear massive gravity with $F(R)=R^n$.}
\end{table}

Next we check if the perturbation around fixed point stays on the
 tangent plane, we compute $
 \vec{n}\cdot\frac{d \delta \vec{x}}{d N}
 $ and find out that for Point $A_1, B_1$ and $C_1$, it automatically equals to
 $0$ suggesting that those points have $0$ eigenvalue, for the rest of the points
 it equals to $0$ if we require that the perturbation satisfy $n \delta x_3  + \delta x_2  = 0 $, which means the perturbation starts on the tangent plane.
 One can see that  points $H$, $I$ in the general case corresponding to $L_1$, $F_1$. we will analyze the stability of each point in detail.

Point$A_1$: The eigenvalues of the linearised system are
\ba
0, -5, 2, \frac{4 n}{n-1}, -3 (\omega +1),\n
\end{align}
corresponding eigenvectors are
\ba
&\left\{ - 1, 0, 1, 0, 0 \right\}, \{  - 1, 0, 0, 1, 0\}, \left\{ { - \frac{{15}}{7}, 0, 0,\frac{8}{7}, 1} \right\}, \n\\
&\left\{ {\frac{5(1-3n)}{6n}, \frac{{ 9n-5}}{{6(n - 1)}},  \frac{{5 - 9n}}{{6n(n - 1)}}, 1, 0} \right\}, \left\{ { - \frac{5}{{3(\omega  + 1)}}, 0, 0, 1, 0} \right\}.\n
\end{align}
The eigenvectors of $-5, 2, \frac{4 n}{n-1}, -3 (\omega +1)$ are normal to $\vec{n}$, so they belong to the tangent plane. If we analyze  the system
with one  variable eliminated we would get the same eigenvalues. The eigenvector of $0$ is not normal to the tangent plane, it is tangent to Line A,  it won't effect the stability
of the point. The point is not stable.

Point $B_1$:  The eigenvalues of the linearised system are
\be
0, -2, -2, \frac{2 n}{n-1}, -3 (\omega +1),\n
\ee
corresponding eigenvectors are£º
\ba
&\{  - 1, 0, 1, 0, 0\}, \{  - 1, 0, 0, 1, 0\},  \{ 0, 0, 0, 0, 0\}, \n\\
&\left\{ {\frac{{n(3n - 2) + 1}}{{n(2n - 1)}}, \frac{2n-1}{{1 - n}},  \frac{2n-1}{{n(n - 1)}}, \frac{{n + 1}}{{1 - 2n}}, 1} \right\}, \left\{ { - \frac{2}{{3(\omega  + 1)}}, 0, 0, 1, 0} \right\}.\n
\end{align}
The eigenvectors of $-2, \frac{2 n}{n-1}, -3 (\omega +1)$   belong to the tangent plane, so they must be the eigenvalues of the system with one variable eliminated, The eigenvectors seem insufficient to determine the base vectors of the tangent plane, so we compute the eigenvalues of the system with one variable eliminated and get  eigenvalues of $-2, -2, \frac{2 n}{n-1}, -3 (\omega +1) $. The eigenvector of $0$ is not normal to the tangent plane, but it won't effect the stability of the point. The point is stable when
\be
0 < n < 1 ,~\omega  >  - 1.\n
\ee
The effective equation of state is
$\omega _{eff}  =  - \frac{1}{3}$.

Point $C_1$: The eigenvalues of the linearised system are
\be
0, -2, \frac{\sqrt{(n-1) (25 n-41)}}{2(1- n)}-\frac{3}{2}, \frac{\sqrt{(n-1) (25 n-41)}}{2 (n-1)}-\frac{3}{2}, -3 (\omega +1),\n
\ee
corresponding eigenvectors are
\ba
 &\{  - 1, 0, 1, 0, 0\}, \left\{ {1-\frac{{2}}{n}, - 2, \frac{2}{n}, 0, 1} \right\}, \n\\
 &\left\{ {\frac{{n -5 +\sqrt {(n - 1)(25n - 41)} }}{{4n}}, \frac{{\sqrt {(n - 1)(25n - 41)} }}{{4(1 - n)}} - \frac{5}{4}, \frac{{ \sqrt {(n - 1)(25n - 41)} }}{{4n(n - 1)}}+\frac{5}{4n}, 1, 0} \right\}, \n\\
 &\left\{ {\frac{{n -5- \sqrt {(n - 1)(25n - 41)} }}{{4n}}, {\frac{{\sqrt {(n - 1)(25n - 41)} }}{{4(n - 1)}} -\frac{5}{4} }, {\frac{{\sqrt {(n - 1)(25n - 41)} }}{{4n(1-n)}} +\frac{5}{4n} }, 1, 0} \right\}, \n\\
 &\left\{ {\frac{{(n - 2)(3\omega(n - 1)  - n - 3)}}{{3n(n - 1)(\omega  + 1)(3\omega  - 1)}}, \frac{2}{{(n - 1)(3\omega  - 1)}}, - \frac{2}{{n(n - 1)(3\omega  - 1)}}, 1, 0} \right\}. \n
\end{align}
Same as Point $A_1, B_1$,  eigenvectors of  the none-zero eigenvalues  belong to the tangent plane therefor  these eigenvalues are responsible for
the stability of the fixed point.
Point $C_1$ is stable when
\ba
1 < n < 2, \omega  >  - 1.
\end{align}
The effective equation of state is
$
\omega _{eff}  =  - 1
$.

Point $D_1$: The eigenvalues of the linearised system are
\be
-2, 2, 2, 2, -1 - 3 \omega,\n
\ee
corresponding eigenvectors are£º
\ba
&\{ 0, 0, 0,  - 1, 1\}, \{ 1, - 2, 0, 0, 1\}, \{  - 1, 0, 0, 1, 0\}, \n\\
&\{  - 1, 0, 1, 0, 0\}, \{ 0, 0, 0, 1, 0\}.
\end{align}
The eigenvectors of $-2, -1 - 3 \omega$ and one of $2 $  belong  to the tangent plane, and the eigenspace of  $2$  intersects with the tangent plane therefor provides  the system with another eigenvalue. The point is not stable.

Point $E_1$: The eigenvalues of the linearised system are
\be
2, 5, 5, \frac{ 5-4n}{1 - n}, 2 - 3\omega,\n
\ee
corresponding eigenvectors are
\ba
&\{ 0, 0, 0, - 1, 1\}, \{  - 1, 0, 0, 1, 0\}, \{  - 1, 0, 1, 0, 0\}, \n\\&\left\{ {0, - \frac{n}{{n - 1}}, \frac{1}{{n - 1}}, 1, 0}\right\}, \{ 0, 0, 0, 1, 0\}.\n
\end{align}
The eigenvectors of $2, \frac{5-4n}{1 - n}, 2 - 3\omega$ and one of $5 $ belong  to the tangent plane, therefor  effect the stability of the point. The point is not stable.

Point $F_1$:The eigenvalues of the linearised system are
\be
2, \frac{4 n-3 (\omega +1)}{n-1}, 3 (\omega +1)~, 3 (\omega +1), 3 \omega -2,\n
\ee
corresponding eigenvectors are
\ba
 &\left\{ {0, 0, 0, \frac{6}{{3\omega  - 4}} + 2, 1} \right\},  \left\{ {0, \frac{{n(3n(\omega  - 2) + 5)}}{{3(n - 1)(\omega(2n - 1)  - 1)}}, - \frac{{3n(\omega  - 2) + 5}}{{3(n - 1)(\omega(2n - 1)  - 1)}}, 1, 0} \right\}, \n\\
&\left\{ { - \frac{5}{{3(\omega  + 1)}}, 0, 0, 1, 0} \right\}, \{  - 1, 0, 1, 0, 0\}, \{ 0, 0, 0, 1, 0\}. \n
\end{align}
The eigenvectors of $2, \frac{4 n-3 (\omega +1)}{n-1}, 3 \omega -2$ and one of $3 (\omega +1) $  belong  to the tangent plane, therefor  effect the stability of the point. The point is not stable.

Point $G_1$:The eigenvalues of the linearised system are
\be
-2, \frac{2 n-3 (\omega +1)}{n-1}, 3 (\omega +1), 3 (\omega +1), 3 \omega +1,\n
\ee
corresponding eigenvectors are
\ba
 &\left\{ {0, 0, 0, \frac{{2\omega }}{{\omega  + 1}}, 1} \right\}, \n\\
 &\left\{ {0, \frac{{ 5- 4n + 3\omega}}{{2(n - 1)}}, \frac{{5 - 4n + 3\omega  }}{{2n(1 - n) }}, - \frac{{(2n - 3(\omega  + 1))(n(6\omega  + 4) - 3\omega  - 5)}}{{2n(n(3\omega  - 1) + 2)}}, 1} \right\}, \n\\
 &\left\{ { - \frac{2}{{3(\omega  + 1)}}, 0, 0, 1, 0} \right\}, \{  - 1, 0, 1, 0, 0\}, \{ 0, 0, 0, 1, 0\}.\n
\end{align}
The eigenvectors of $-2, \frac{2 n-3 (\omega +1)}{n-1}, 3 \omega +1$ and one of $3 (\omega +1) $   belong  to the tangent plane, therefor  effect the stability of the point. The point is stable when
\be
{n < 0,~ \omega  < \frac{1}{3}(2n - 3)}\n~~ or~~
0 < n < 1,~ \omega  <  - 1.\n
\ee
The effective equation of state is  $\omega _{eff}  =  - \frac{1}{3}$.

Point $J_1$: The eigenvalues of the linearised system are
\be
2 n, 2 n, n- \sqrt{3n (3 n-4)}-2, n+ \sqrt{3n (3 n-4)}-2, 2 n-3 (\omega +1),\n
\ee
corresponding eigenvectors are
\ba
 &\left\{ {\frac{1}{{2n(n - 1) - 1}}, \frac{n}{{2n(n - 1) - 1}} - 1, 0, \frac{{n + 1}}{{1 - 2n(n - 1)}}, 1} \right\}, \{  - 1, 0, 1, 0, 0\}, \n\\
 &\left\{ {0, \frac{{n(5 - 4n) -  \sqrt {3n(3n - 4)} }}{{2(2n(n - 1) - 1)}}, -\frac{{n(5 - 4n) -  \sqrt {3n(3n - 4)} }}{{2n(2n(n - 1) - 1)}}, \frac{{n\left( {7 - 5n } \right) + (n-1) \sqrt {3n(3n - 4)} }}{{2n(2n(n - 1) - 1)}}, 1} \right\}, \n\\
 &\left\{ {0, \frac{{n(5 - 4n) + \sqrt {3n(3n - 4)} }}{{2(2n(n - 1) - 1)}}, -\frac{{ n(5 - 4n) + \sqrt {3n(3n - 4)} }}{{2n( 2n(n-1  )-1)}}, \frac{{ (1-n)\sqrt {3n(3n - 4)}  - n\left( {5n  - 7} \right)}}{{2n(2n(n - 1) - 1)}}, 1} \right\}, \n\\
 &\left\{ {0, \frac{{2n(2n-3)+ 3\omega  + 1}}{{2(1 - 2n(n - 1))}}, - \frac{{2n(2n-3) + 3\omega  + 1}}{{ 2n(1 - 2n(n-1  ))}}, - \frac{{(2n - 3\omega  - 1)(2n - 3(\omega  + 1))}}{{4n(2n(n - 1) - 1)}}, 1} \right\}.\n
\end{align}
The eigenvectors of $n- \sqrt{3n (3 n-4)}-2, n+ \sqrt{3n (3 n-4)}-2, 2 n-3 (\omega +1)$  belong to the tangent plane, and the eigenspace of $2n$ intersects with
tangent plane, therefor they all effect the stability of the point. The point is stable when
\be
\frac{1}{2} - \frac{{\sqrt 3 }}{2} < n < 0,~
\omega  > \frac{1}{3}(2n - 3).\n
\ee
The effective equation of state is $\omega _{eff}  =  - \frac{1}{3}$.

Point $K_1$:  The eigenvalues of the linearised system are
\ba
&\frac{1}{n-1}-4,~-\frac{2 n(n-2) }{(n-1) (2 n-1)},~-\frac{2n (n-2) }{(n-1) (2 n-1)},~\n\\
&\frac{2n-4}{(1-2 n)(n-1)}-2,~-\frac{1+n}{(n-1)(1-2n)}-3 \omega -4,\n
\end{align}
corresponding eigenvectors are£º
\ba
 &\left\{ {0, \frac{{n \left(8 n^2-22 n+15\right)}}{{(2n(2n - 5) + 7)(1-n)}} , \frac{{2n-1}}{{(2n(2n - 5) + 7)(1-n)}}, \frac{{4(n - 2)(n - 1)}}{{2n(2n - 5) + 7}}, 1} \right\}, \n\\
 &\left\{ {0, - \frac{n}{{n - 1}}, \frac{1}{{n - 1}}, 1, 0} \right\}, \left\{ {\frac{{2(8 - 3n)n - 11}}{{6(n - 1)^2 }}, \frac{{5 - 4n}}{{6(n - 1)^2 }}, 0, 1, 0} \right\}, \{- 1, 0, 1, 0, 0\}, \n\\
 &\left\{ {0, \frac{{n(4n - 5)}}{{3(n - 1)^2 ((2n - 1)\omega  - 1)}}, \frac{{5 - 4n}}{{3(n - 1)^2 ((2n - 1)\omega  - 1)}}, 1, 0} \right\}.\n
\end{align}
None of the eigenvectors  are parallel  to  the normal vector  therefor all the eigenvalues  effect the stability of the point, however when $n$ equals $\frac{5}{4}$ or $\frac{3}{2}$, eigenvector of $\frac{1}{n-1}-4$ may be parallel to $\vec{n}$, but in that case the point is not stable. The point is stable when
\be
n < \frac{1}{2}\left( {1 - \sqrt 3 }\right),  or~ \frac{1}{2} < n < 1, or~ n > 2\n
\ee
and
\be
{ \omega  > \frac{{ - 8n^2  + 13n - 3}}{{6n^2  - 9n + 3}}}.\n
\ee
The effective equation of state is
$\omega _{eff}  = \frac{{ n(7- 6n ) + 1}}{{6n^2  - 9n + 3}}$.

Point $L_1$:  The eigenvalues of the linearised system are
\ba
&3 (\omega +1), 3 (\omega +1), \frac{-2 n+3 \omega +3}{n}, \n\\
&\frac{-\sqrt{n-1} \sqrt{4 n^3 (3 \omega +8)^2+\cdots}+3 n ((2 n-3) \omega -1)+3 \omega +3}{4 (n-1) n}, \n\\
&\frac{\sqrt{n-1} \sqrt{4 n^3 (3 \omega +8)^2+\cdots}+3 n ((2 n-3) \omega -1)+3 \omega +3}{4 (n-1) n}. \n
\end{align}
Both the  specific forms of  corresponding eigenvectors and the parameter range for the point to be stable  are complicated,  we do not give their exact forms, the point can be stable. The effective equation of state is $
\omega _{eff}  = \frac{{ 1 + \omega }}{n}-1$, the point is  a solution depending on matter fluid with $\Omega _m  = \frac{{n(9\omega  + 13 - 2n (3\omega  + 4))   - 3(\omega  + 1)}}{{2n^2 }}$.

\subsection{$ln(R)$ model}
Let us discuss now the case of  Lagrangian $
F(R)= ln (R).
$
Corresponding constraint surface can be  written as
\be
x_1  - \gamma x_2  = 0\left( {\gamma  = \frac{{2\rho _{ir} }}{{M_p^2 }}} \right).
\ee
The normal vector is $\left( {1, - \gamma, 0, 0, 0} \right)$.
First we check if the surface is an integral, after some calculation we get
\be
\vec{n}\cdot\frac{d\vec{x}}{dN}=(x_1-\gamma  x_2) (-2( x_2+x_5)+x_4+4),
\ee
so the surface is an  integral.
$m$ takes the constant value of $-1$. We summarize the fixed points of this autonomous system and their stability analysis in
Table III.

\begin{table}
\begin{tabular}{ccccccccc}

\hline
Lable &${x_1}$ & ${x_2}$ & ${x_3}$ & ${x_4}$ & ${x_5}$ & ${\Omega_m}$ & $\omega_{eff}$ & Stable when \\
\hline$A_2$&0&0&5&-4&0&0& $\frac{1}{3}$ & \\
$B_2$&0&0&2&-2&1&0& $-\frac{1}{3}$ &  \\
$C_2$&$2\gamma$ &2& $- 2\gamma  - 1$&0&0&0& $-1$& Not stable \\
$D_2$&0&0&0&0&1&0& $-\frac{1}{3} $ &  Not stable\\
$E_2$&0&0&0&1&0&0&$\frac{1}{3} $ & Not stable \\
$F_2$&0&0&0&$- 1 + 3\omega $&0&$2-3 \omega$& $\frac{1}{3} $ & Not stable \\
$G_2$&0&0&0&$1 + 3\omega$&1&$-3 \omega -1$& $-\frac{1}{3} $ & $
\omega  <  - 1$\\
\hline

\end{tabular}
\caption{The fixed points  in F(R) nonlinear massive gravity with
$F(R) = ln R$}
\end{table}

Next we check if  the perturbation stays on the
 tangent plane at fixed point. We compute
 $
\vec{n}\cdot\frac{d \delta \vec{x}}{d N}
 $ and find out that for Point $ A_2, B_2$ and $C_2$, it  automatically equals to
 $0$ suggesting that those points have $0$ eigenvalue, for the rest of the points
 it equals to $0$ if we require that the perturbation satisfy $\delta x_1  - \gamma \delta x_2  = 0$, which means the perturbation starts on the tangent plane.
We now analyze each fixed point in detail.

Point $A_2 $: The eigenvalues of the linearised system are
\be
-5, 2, 0, 0, -3 (\omega +1), \n
\ee
corresponding eigenvectors are
\ba
 \{ 0, 0, - 1, 1, 0\}, \left\{ {0, 0,  - \frac{{15}}{7}, \frac{8}{7}, 1} \right\}, \left\{ { - \frac{{11}}{6}, \frac{5}{6}, 0, 1, 0} \right\}, \n\\
 \{  - 1, 0, 1, 0, 0\}, \left\{ {0, 0,  - \frac{5}{{3(\omega  + 1)}}, 1, 0} \right\}.\n
\end{align}
The eigenvectors of $-5, 2, -3 (\omega +1)$  are normal to $\vec{n}$, so they
belong to the tangent plane, therefor  $-5, 2, -3 (\omega +1)$ are eigenvalues of the system with the redundant variable eliminated. The eigenspace
of $0$ intersects with  tangent plane, so $ 0$ is also the eigenvalues of the
system with the redundant variable eliminated. The stability
of this doubly degenerate equilibrium can be analysed with more advance
technique, and is beyond the   scope of this paper.

 Point $B_2 $: The eigenvalues of the linearised system are
\be
-2, -2, 0, 0, -3 (\omega +1),\n
\ee
corresponding eigenvectors are
\ba
&\{0, 0, -1, 1, 0\}, \{0, 0, 0, 0, 0\}, \{-1, -1, 0, 1, 1\}, \n\\
&\{-1, 0, 1, 0, 0\}, \left\{0, 0, -\frac{2}{3 (\omega +1)}, 1, 0\right\}.\n
\end{align}
Therefor  $-2, -3 (\omega +1), 0$ are eigenvalues of the system with the redundant variable eliminated. Point $B_2$ is also   doubly degenerate equilibrium same as Point $A_2$, we will not discuss its stability.

Point $C_2$: The eigenvalues of the linearised system are
\be
-\frac{1}{2} \left(3+\sqrt{41}\right), -2, \frac{1}{2} \left(\sqrt{41}-3\right), 0, -3 (\omega +1),\n
\ee
corresponding eigenvectors are
\ba
&\left\{-\frac{1}{4} \left(5+\sqrt{41}\right) \gamma , \frac{4}{5-\sqrt{41}}, \frac{1}{4} \left(\left(5+\sqrt{41}\right) \gamma +\sqrt{41}+1\right), 1, 0\right\}, \n\\
&\{-2 \gamma , -2, 2 \gamma +1, 0, 1\},
\left\{\frac{4 \gamma }{5+\sqrt{41}}, \frac{1}{4} \left(\sqrt{41}-5\right), \frac{1}{4} \left(\left(5-\sqrt{41}\right) \gamma -\sqrt{41}+1\right), 1, 0\right\},\n\\
&\{-1, 0, 1, 0, 0\}, \left\{\frac{2 \gamma }{1-3 \omega }, \frac{2}{1-3 \omega }, \frac{6 \gamma  (\omega +1)+9 \omega +1}{9 \omega ^2+6 \omega -3}, 1, 0\right\}.\n
\end{align}
Here the eigenvector of $0$ is tangent to Line C, we are able to analyze the stability of this point. None of the eigenvectors  are parallel to $\vec{n}$, so they all are responsible for the stability of the fixed point(eigenvector of $\frac{1}{2} \left(\sqrt{41}-3\right)$ may be parallel to $\vec{n}$, but it requires $\gamma$ to be imaginary). Because $\frac{1}{2} \left(\sqrt{41}-3\right)$ is  positive, the point is not stable.

Point $D_2$:
The eigenvalues of the linearised system are
\be
-2, 2, 2, 2, -3 \omega -1\n
\ee
Corresponding eigenvectors are
\ba
&\{0, 0, 0, -1, 1\}, \{1, -2, 0, 0, 1\}, \{-1, 0, 0, 1, 0\}, \n\\
&\{-1, 0, 1, 0, 0\}, \{0, 0, 0, 1, 0\}.\n
\end{align}
Eigenvectors of $ -2, -3 \omega -1$ are on the tangent plane,  The eigenspace
of $2$  intersects with  tangent plane, so the system  with the redundant variable eliminated  have the eigenvalues of $ 2, 2, -2, -3 \omega -1$.
The point is not stable.

Point $E_2$: The eigenvalues of the linearised system are
\be
5, 5, 5, 2, 2-3 \omega,\n
\ee
corresponding eigenvectors are
\ba
&\{-1, 0, 0, 1, 0\}, \{-1, 0, 1, 0, 0\}, \{0, 0, 0, 0, 0\}, \n\\
&\{0, 0, 0, -1, 1\}, \{0, 0, 0, 1, 0\}.\n
\end{align}
Eigenvectors of $ 2, 2-3 \omega$ are on the tangent plane,   The eigenspace
of $5$   intersects with  tangent plane, so the system with  the redundant variable eliminated  have the eigenvalues of $ 5, 5, 2, 2-3 \omega$ (the eigenspace of $5 $ seems insufficient to determine a base vector of the tangent plane, so we compute the eigenvalues of the system with one variable eliminated and get the desired eigenvalues).
The point is not stable.

Point $F_2$:
The eigenvalues of the linearised system are
\be
2, 3 (\omega +1), 3 (\omega +1), 3 (\omega +1), 3 \omega -2, \n
\ee
corresponding eigenvectors are
\ba
&\left\{0, 0, 0, \frac{6}{3 \omega -4}+2, 1\right\}, \left\{-\frac{5}{3 (\omega +1)}, 0, 0, 1, 0\right\}, \{-1, 0, 1, 0, 0\}, \n\\
&\{0, 0, 0, 0, 0\}, \{0, 0, 0, 1, 0\}.\n
\end{align}
eigenvectors of $ 2, 3 \omega -2$ are on the tangent plane. The eigenspace
of $3 (\omega +1)$   intersects with  tangent plane, so the system with   the redundant variable eliminated  have the eigenvalues of $ 2, 3 \omega -2, 3 (\omega +1), 3 (\omega +1)$ (the same situation as Point $E_2$).
The point is not stable.

Point $G_2$:
The eigenvalues of the linearised system are
\be
-2, 3 (\omega +1), 3 (\omega +1), 3 (\omega +1), 3 \omega +1\n
\ee
Corresponding eigenvectors are
\ba
&\left\{0, 0, 0, \frac{2 \omega }{\omega +1}, 1\right\}, \left\{-\frac{2}{3 (\omega +1)}, 0, 0, 1, 0\right\}, \{-1, 0, 1, 0, 0\}, \n\\
&\{0, 0, 0, 0, 0\}, \{0, 0, 0, 1, 0\}.\n
\end{align}
Eigenvectors of $ -2, 3 \omega +1$ are on the tangent plane, The eigenspace
of $3 (\omega +1)$   intersects with  tangent plane, so the system   with the redundant variable eliminated  have the eigenvalues of $3 (\omega +1), 3 (\omega +1), -2, 3 \omega +1 $ (the same situation as Point $E_2$). The point is stable when $\omega<-1$, the effective equation of state is $\omega _{eff}  =  - \frac{1}{3}$.

\section{Conclusions}

In this paper we have studied the dynamical behavior of the $F(R)$
nonlinear massive gravity by recasting the field equations   into
a $6$ dimensional autonomous system. However after  reducing  the
dimension of the  system by one via  the constraint
equation, we could see that the system still has a redundant variable which  gives rise to a hidden
constraint equation. The hidden
constraint equation which  depends on the  model of $F(R)$
would change the behavior of the perturbation greatly, and ignoring it may result in false conclusion in the
stability analysis of the fixed points.  We study the
stability of the fixed points by analyzing the relations between
the eigenvector of a certain eigenvalue and the tangent plane of
the  constraint surface at the fixed point. If the eigenvector is
normal to the tangent plane, the corresponding eigenvalue do not
effect the stability of the fixed point, otherwise the eigenvalue
should be considered. We analyze the system in this way instead
of eliminating one variable because it would  result in
complicated relations between variables thus making the dynamic of
the  system utterly complicated and hard to analyze.

Notice that  some lines of equilibria have emerged instead of
fixed points, this situation happens when the line of equilibria
has a $0$ eigenvalue whose eigenvector is tangent to the line of
equilibria. The  lines  of equilibria would intersect with the
constraint surface and the points of intersection are considered as
the fixed points of the system, and we could carry out the
analysis as discussed above.

We consider two specific models  of $F(R)$ which are $R^n$ and $ln(R)$. The models are
relatively simple and the  specific  forms of constraint surfaces are easy to obtain, but the same process can be carried out with a more complicated
 model  of $F(R)$. Both models present a few stable points which may have
interesting cosmological implication. However $F(R)$ nonlinear
massive gravity possesses plentiful phenomenological properties
due to its features inherited both from nonlinear massive gravity
and $F(R)$ gravity. More study is needed to fully understand the
cosmological behavior of this model.

\textbf{Acknowledgments} The author thanks Yi-Fu Cai and Yun-Song
Piao for helpful discussions and comments. This work is supported
in part by NSFC under Grant No:11222546, in part by National Basic
Research Program of China, No:2010CB832804.

\end{document}